\documentclass[journal]{IEEEtran}
 
\usepackage{graphicx}

\newcommand{\samefig}[1]{%
  \includegraphics[width=\linewidth,height=0.72\linewidth]{#1}%
}
\usepackage{float}
\usepackage{caption}
\usepackage{subcaption} 
\usepackage[ruled,vlined]{algorithm2e} 
\usepackage{amsmath} 
\usepackage{amsthm,amssymb}
\usepackage{amsfonts}
\usepackage{bm}
\usepackage{dsfont}
\usepackage{xcolor}
\usepackage{xurl}
\usepackage{stmaryrd}
\usepackage{booktabs} 
\usepackage{array,booktabs,multirow,makecell}
\usepackage{tabularx}
\usepackage{capt-of}
\usepackage{needspace}
\usepackage{booktabs}
\usepackage{subcaption}
\usepackage{fancyhdr}
\usepackage{color}
\usepackage{ragged2e}
\usepackage{balance}
\usepackage{xspace}
\usepackage{mathtools}
\newsavebox{\figabox}
\usepackage[colorlinks=true]{hyperref}
\hypersetup{
    linkcolor = {red},
    citecolor = {black},
    urlcolor = {cyan},
}

\begin{document}
\title{Low-Altitude Fluid Antenna Network with Multi-Agent Reinforcement Learning}

\author{Tong Zhang,  \textit{IEEE Member}, Yanfei Su, Shuai Wang,  \textit{IEEE Senior Member}, Wanli Ni, \textit{IEEE Member}, \\ Chengzhong Xu, \textit{IEEE Fellow},  and Hüseyin Arslan, \textit{IEEE Fellow}


\thanks{T. Zhang is with Guangdong Provincial Key Laboratory of Aerospace Communication and Networking Technology, Harbin Institute of Technology, Shenzhen, 518055, China (e-mail: tongzhang@hit.edu.cn).}
\thanks{Y. Su is with Harbin Institute of Technology, Shenzhen, China (e-mail: 2022210389@stu.hit.edu.cn).}
\thanks{S. Wang is with the Shenzhen Institutes of Advanced Technology, Chinese Academy of Sciences, Shenzhen 518055, China (e-mail: s.wang@siat.ac.cn).}
\thanks{W. Ni is with the School of Information and Communication Engineering, Beijing University of Posts and Telecommunications, Beijing 100876, China (e-mail: niwanli@bupt.edu.cn).}  
\thanks{C. Xu is with State Key Laboratory of Internet of Things for Smart City, University of Macau, Macau SAR 999078, China (e-mail: czxu@um.edu.mo).}
\thanks{H. Arslan is with the Department of Electrical and Electronics Engineering, Istanbul Medipol University, Istanbul 34810, Türkiye (e-mail: huseyinarslan@medipol.edu.tr).}
\thanks{Tong Zhang and Yanfei Su contributed equally to this work.}
\thanks{Corresponding author: Shuai Wang.}
}
\maketitle

\begin{abstract}
Low-altitude wireless networks (LAWNs) integrate terrestrial and aerial platforms to provide ubiquitous communication, sensing, and localization services for unmanned aerial vehicles (UAVs) and electric vertical takeoff and landing (eVTOL) aircrafts.  However, dynamic air-ground and air-air channels, abrupt blockages, and heterogeneous interference hinder the realization of this goal. Nevertheless, fluid antenna (FA), a cutting-edge multiple-input multiple-output (MIMO) technique, overcomes these challenges by reconfiguring antenna positions to unlock additional spatial degrees-of-freedom. In this paper,  towards bringing low-altitude FA networks into reality, we study the fast and high-performance FA reconfiguration for low-altitude FA networks with multi-agent reinforcement learning (MARL). Specifically, we present an electromagnetic digital twin (EM-DT)-assisted MARL framework. To fill the sim-to-real gap, we introduce a two-stage transfer learning framework. Our case study shows that joint FA positions and beamforming optimization can enhance the system sum-rate by $118.5\%$, compared to the fixed-position baseline. This gain comes from the dynamic millisecond-timescale reconfiguration of FA arrays and the adaptive steering of beams toward aerial users with mobility.
\end{abstract}

\section{Introduction}

\IEEEPARstart{L}{ow-altitude} economy refers to economic activities based on the regular operation of manned and unmanned aircraft in low-altitude airspace, typically below 1,000 meters and up to 3,000 meters \cite{zhao2026generative}. These activities span package delivery, agricultural monitoring, infrastructure inspection, emergency response, and future air taxi services, etc. To support these diverse use cases, wireless networks are required to provide ubiquitous communication, sensing, and localization services for both unmanned aerial vehicles (UAVs) and electric vertical takeoff and landing (eVTOL) aircraft, a capability beyond the reach of current terrestrial networks. Specifically, terrestrial networks struggle with frequent blockage-induced line-of-sight (LoS)/non-line-of-sight (NLoS) transitions, mobility-induced rapid channel variations, and severe inter-cell interference from aerial LoS connectivity to multiple BSs in three-dimensional (3D) airspace. To address these limitations, the low-altitude wireless network (LAWN) framework has been proposed to integrate terrestrial base stations (BSs) with aerial users, with the goal of providing ubiquitous communication, pervasive sensing and localization, as well as high-data-rate, low-latency, and ultra-reliable connectivity \cite{jun2026low}.

  \begin{figure*}[t]
    \centering

    \sbox{\figabox}{\includegraphics[width=0.64\textwidth]{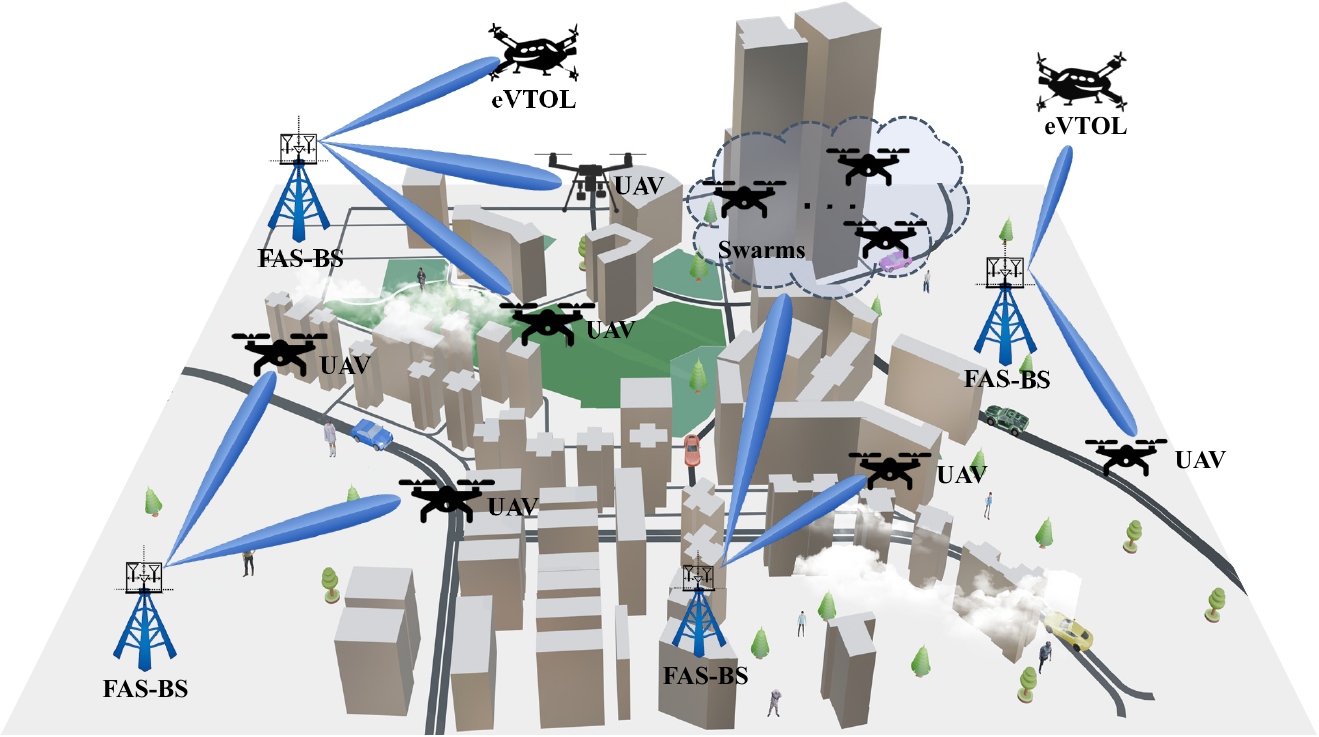}}%

    \begin{subfigure}[t]{0.64\textwidth}
        \centering
        \vspace{0pt}
        \newlength{\figAwidth}
        \setlength{\figAwidth}{1.15\linewidth}

        \makebox[\linewidth][l]{%
             \includegraphics[
                width=1.15\linewidth,
                height=\ht\figabox,
                keepaspectratio=false
            ]{F1.pdf}%
        }
       \makebox[\linewidth][l]{%
        \begin{minipage}{1.15\linewidth}
            \centering
            \subcaption{Low-altitude FA network.}
            \label{figf1network}
        \end{minipage}%
        }
    \end{subfigure}
    \hfill 
    \begin{subfigure}[t]{0.31\textwidth}
        \centering
        \vspace{0pt}
        \includegraphics[width=0.58\linewidth]{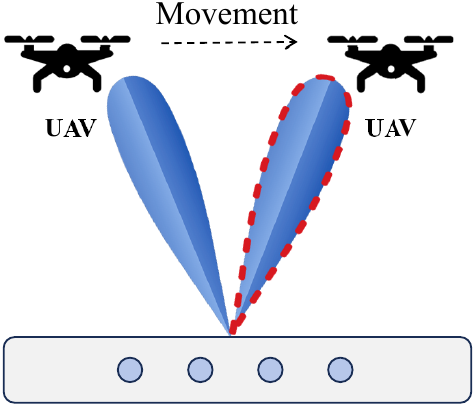}
        \subcaption{Traditional MIMO.}
        \label{fig:f1_cmimo} \vspace{0.5cm}
        \includegraphics[width=0.64\linewidth]{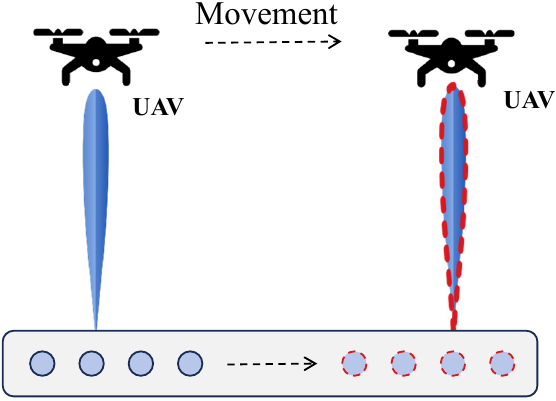}
        \subcaption{FA MIMO.}
        \label{fig:f1_famimo}
    \end{subfigure}
    \caption{Illustration of a low-altitude FA network (Subfig. {\color{red}{a}}), and conceptual comparison between traditional MIMO and FA MIMO (Subfigs. \subref{fig:f1_cmimo} and \subref{fig:f1_famimo}).}
    \label{F1}
\end{figure*}

However, the mobility of aerial users gives rise to dynamic air‑ground and air‑air channels, sudden blockages, and heterogeneous interference. These difficulties are unlikely to be resolved by traditional MIMO systems, since their fixed arrays are ill‑suited to tracking fast‑moving UAVs and eVTOLs.

Fluid antenna (FA), which is a cutting-edge MIMO technology, offers an additional spatial degree of freedom by reconfiguring antenna positions, shapes, and even radiation patterns \cite{Wu2026FAS,Wu2026SFAS,Yang2025FASISAC,zhang2026indoor}.  Unlike traditional MIMO beamforming, which mainly changes the excitation weights of a fixed array, FAs also reshape the array geometry and change to more favorable antenna positions. This position-domain controllability makes FAs especially valuable for LAWNs, where the trajectory-dependent variations in line-of-sight (LoS) links, blockages, and heterogeneous interference render traditional antennas insufficient. Therefore, low-altitude FA networks have been proposed to upgrade LAWNs by leveraging FA capabilities \cite{liu2026fluid, Zuo2026FA,Guo2026FAS,Zhang2026UAV}.  
In \cite{liu2026fluid}, the authors showed that FA systems can greatly improve communication, sensing, and control in LAWNs by flexibly adjusting antenna positions. In \cite{Zuo2026FA}, the authors proposed an FA-empowered integrated sensing and communication (ISAC) system for low-altitude economy networks to enhance both communication and sensing performance. In \cite{Guo2026FAS}, the authors proposed a  FA-empowered anti-jamming communication architecture for LAWNs to maximize the worst-case achievable rate under malicious jamming attacks. In \cite{Zhang2026UAV}, the authors proposed a UAV-enabled FA system for LAWNs, jointly optimizing trajectory, antenna positions, and beamforming to minimize sensing error and achieve high-accuracy multi-target sensing. To bring low-altitude FA networks into reality, we focus on how electromagnetic digital twin (EM-DT),  multi-agent reinforcement learning (MARL), and transfer learning can jointly provide high network utility under high-mobility, heterogeneous interference, and limited real-world measurements. This raises the following question: \textit{Can we rapidly reconfigure FA positions to maximize network utility and adapt to the high-mobility of UAVs and eVTOLs?}

In this paper, we attempt to answer this question by making the following main contributions:
\begin{itemize}
    \item We envision a low-altitude FA network  with a unified view, and propose a novel synthesis of EM-DT, MARL, and sim-to-real transfer learning to enable robust operation of this low-altitude FA network.
    \item We show that joint optimization of FA positions and downlink beamforming can yield a substantial 118.5\% sum-rate gain over traditional fixed-position baselines, a result enabled by millisecond-timescale FA reconfiguration and agile beam steering that tracks aerial users.
    \item We discuss key unresolved issues and chart potential avenues for future MARL-driven low-altitude FA networks.
\end{itemize}

\section{Low-Altitude Fluid Antenna Network}

We envision a low-altitude FA network with the architecture and core functionalities specified below. Distinct from the existing frameworks in \cite{liu2026fluid, Zuo2026FA, Guo2026FAS, Zhang2026UAV}, our perspective offers a unified view that  integrates communication, sensing, and localization via FA positioning across both users and BSs.
 
\subsection{System Architecture}

The physical entities of low-altitude FA networks can be categorized into two major entities: aerial users and ground/aerial BSs. Fig.~\ref{figf1network} illustrates an exemplified low-altitude FA network, where aerial users are served by FAS-BSs. Figs.~\ref{fig:f1_cmimo} and \ref{fig:f1_famimo} further illustrate the key distinction between traditional fixed-array beamforming and joint FA positioning and beamforming, where the latter one provides more accurate beamforming due to changing of FA positions.

\textit{1) Aerial Users:}  Aerial users can be enhanced by FAS. 
FA-empowered aerial users need high-data-rate, low-latency, and ultra-reliable uplink/downlink connectivity, as well as ubiquitous communications. They can also support aerial user-driven distributed sensing and aerial user-to-user communication. 

\textit{2) BSs:}  
Both ground BSs and aerial BSs can benefit from FAS. Ground FAS-BSs can provide high-quality commercial services such as enhanced mobile broadband for aerial users and prompt emergency services like reliable links for first responders, while also enabling pervasive sensing. Moreover, they dynamically adjust their FA positions to shift beams toward aerial users, mitigate interference through spatial reconfiguration, and support seamless handover when aerial users move across cell boundaries. On the other hand, aerial FAS-BSs can serve as on‑demand flying BSs or relays to extend coverage, offload terrestrial BSs, and provide connectivity in remote or disaster‑stricken areas. By reconfiguring their FA positions, they can dynamically steer beams toward ground BSs and other aerial users/BSs, suppress co‑channel interference, and combat dynamic air-ground and air-air channels.

Furthermore, these physical entities, including users and BSs, interact through the control domain. Specifically, each BS coordinates user handover and interference information in the control domain. On the other hand, aerial users report channel state information (CSI), positions, and kinematics to the control domain. All of them collectively enable real-time adaptation to dynamic channel conditions and management of user mobility and heterogeneous interference.

\begin{figure*}[t]
\centering

\begin{subfigure}{0.235\textwidth}
  \centering
  \includegraphics[width=\linewidth]{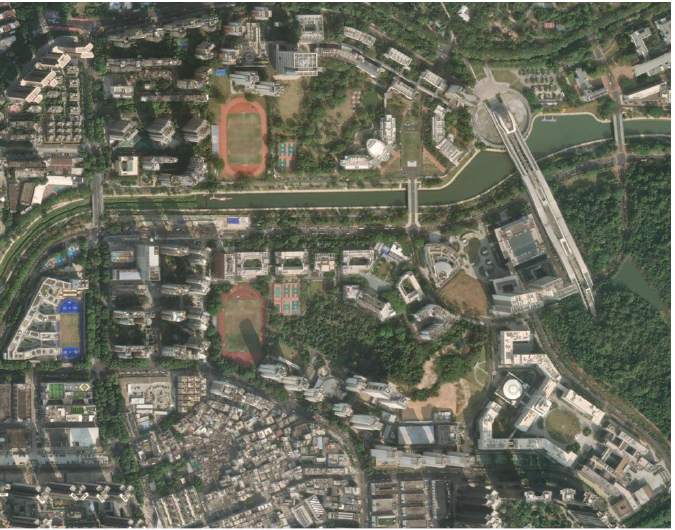}
  \subcaption{}\label{fig:dt_satellite_a}
\end{subfigure}\hspace{0.008\textwidth}
\begin{subfigure}{0.235\textwidth}
  \centering
  \includegraphics[width=\linewidth]{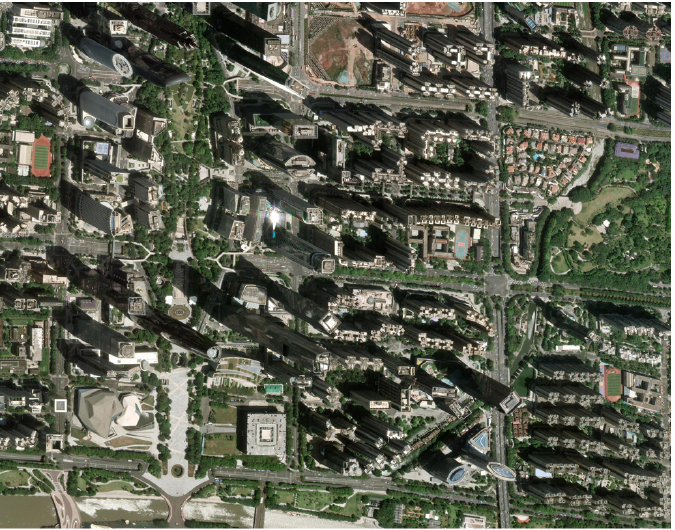}
  \subcaption{}\label{fig:dt_satellite_b}
\end{subfigure}\hspace{0.008\textwidth}
\begin{subfigure}{0.235\textwidth}
  \centering
  \includegraphics[width=\linewidth]{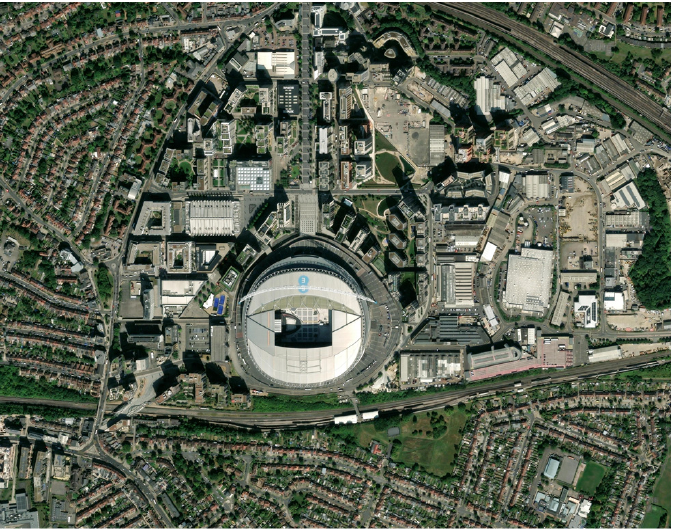}
  \subcaption{}\label{fig:dt_satellite_c}
\end{subfigure}\hspace{0.008\textwidth}
\begin{subfigure}{0.235\textwidth}
  \centering
  \includegraphics[width=\linewidth]{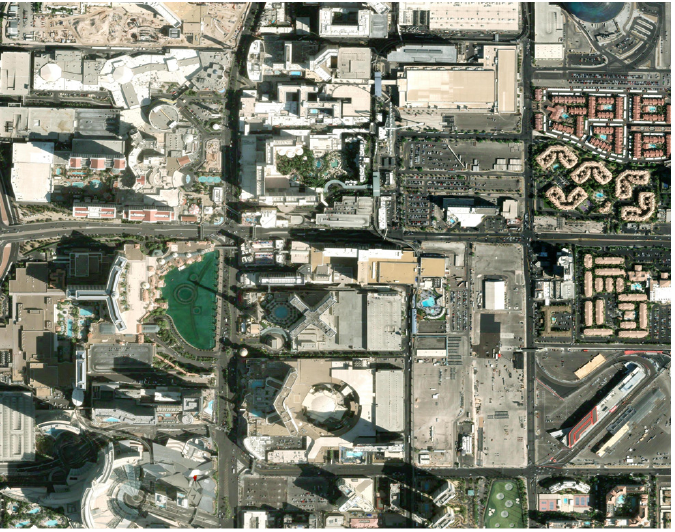}
  \subcaption{}\label{fig:dt_satellite_d}
\end{subfigure}

\vspace{2mm}

\begin{subfigure}{0.235\textwidth}
  \centering
  \includegraphics[width=\linewidth]{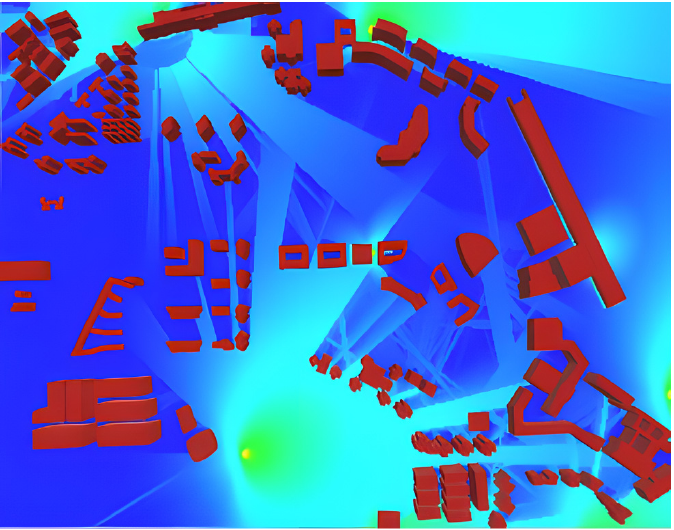}
  \subcaption{}\label{fig:dt_satellite_e}
\end{subfigure}\hspace{0.008\textwidth}
\begin{subfigure}{0.235\textwidth}
  \centering
  \includegraphics[width=\linewidth]{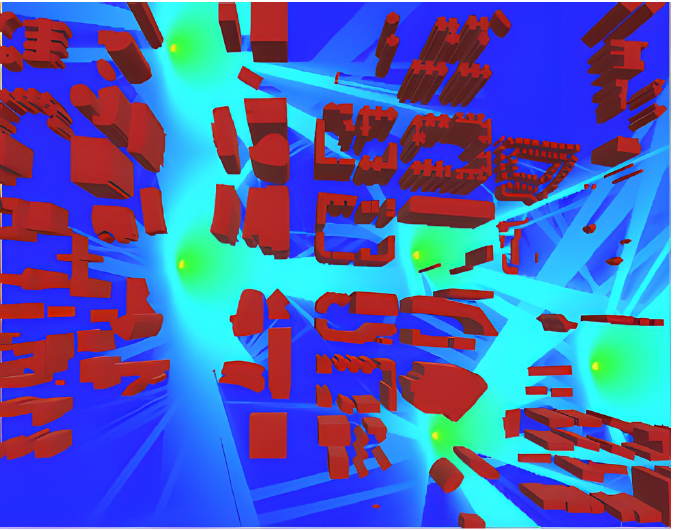}
  \subcaption{}\label{fig:dt_satellite_f}
\end{subfigure}\hspace{0.008\textwidth}
\begin{subfigure}{0.235\textwidth}
  \centering
  \includegraphics[width=\linewidth]{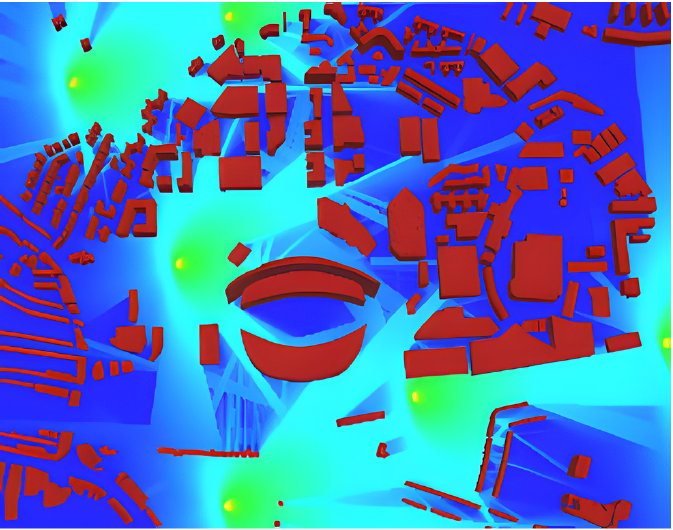}
  \subcaption{}\label{fig:dt_satellite_g}
\end{subfigure}\hspace{0.008\textwidth}
\begin{subfigure}{0.235\textwidth}
  \centering
  \includegraphics[width=\linewidth]{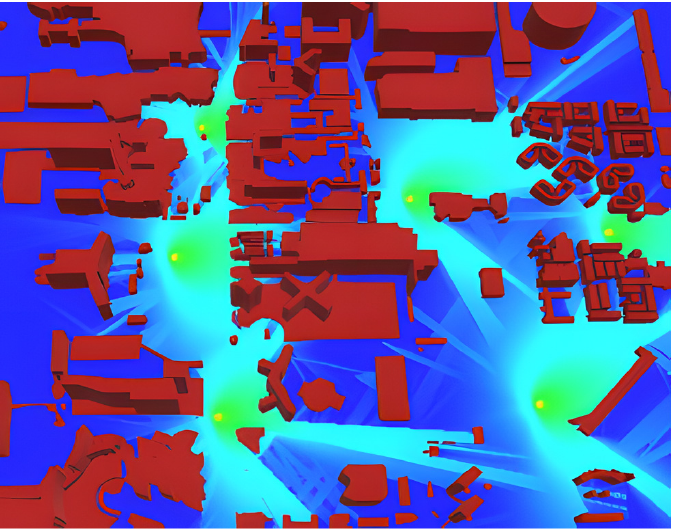}
  \subcaption{}\label{fig:dt_satellite_h}
\end{subfigure}

\caption{Representative EM-DTs and their satellite imagery. The top row shows the satellite imagery and the bottom row shows the corresponding EM-DTs created by \textsc{RANPLAN ACADEMIC V7.1} software, with Shenzhen University Town, China (Subfigs. \subref{fig:dt_satellite_a} and \subref{fig:dt_satellite_e}), Guangzhou Zhujiang New Town CBD, China (Subfigs. \subref{fig:dt_satellite_b} and \subref{fig:dt_satellite_f}), Wembley Stadium in London, UK  (Subfigs. \subref{fig:dt_satellite_c} and \subref{fig:dt_satellite_g}), and Las Vegas Strip, USA (Subfigs. \subref{fig:dt_satellite_d} and \subref{fig:dt_satellite_h}).}
\label{fig:dt_satellite_scenarios}
\end{figure*}
\subsection{Key Functionalities}

Low-altitude FA networks are expected to provide the following main functionalities required by low-altitude economy applications, such as package delivery, agricultural monitoring, infrastructure inspection,  and future air taxis.  Compared with LAWNs, these functionalities, listed below, can be more readily satisfied through the use of FAs, since FA positions can be reconfigured to achieve more favorable channel conditions.

\textit{1) Ubiquitous Communication:} Ubiquitous wireless connectivity is essential for low-altitude economy applications. Low-altitude FA networks should extend connectivity from terrestrial users to aerial users, particularly in urban canyons, remote rural areas, and high-mobility air corridors.

\textit{2) Pervasive Sensing and Localization:} To enable UAV and eVTOL tracking, collision avoidance, and airspace monitoring across the low-altitude airspace, low-altitude FA networks should provide both pervasive sensing and localization. To enable reliable sensing and localization, the low-altitude FA networks could integrate  multiple measurements such as time difference of arrival, angle of arrival, and received signal strength from BSs and aerial users. All sensing and localization data can be fused to form a unified situational view, thereby enhancing navigation safety and airspace awareness.

\textit{3) High-Data-Rate, Low-Latency, and Ultra-Reliable Communication:} To fulfill low-altitude economy application, low-altitude FA networks must accommodate heterogeneous quality of service (QoS) requirements, including high-throughput data delivery, low-latency control signaling, and ultra-reliable communication for mission-critical aerial operations.

\section{Electromagnetic Digital Twin-Assisted Multi-Agent Reinforcement Learning}

\subsection{Electromagnetic Digital Twin}
In low-altitude FA networks, the major challenge is that obtaining training samples from real systems is expensive, risky, and often lacks ground-truth labels. Hence, EM-DT is introduced to resolve this issue. EM-DT serves as a high-fidelity virtual replica of the real low-altitude airspace via modeling electromagnetic wave properties \cite{wang2025electromagnetic}. Beyond offline channel simulation, it couples propagation, mobility, interference, and FA control actions in a closed loop, allowing a controller to test ``what-if'' position and beamforming updates before real deployment. As such, EM-DT enables efficient, scalable, and risk-free MARL policy training and evaluation. Four EM-DTs in representative areas are shown in Fig.~\ref{fig:dt_satellite_scenarios}.

Building an EM-DT requires accurately modeling three core cornerstones. The first cornerstone is propagation modeling, which relies on 3D ray-tracing techniques to capture diffraction, reflection, and scattering effects specific to low-altitude airspace. The second cornerstone is user mobility dynamics modeling, including the 3D trajectories of UAVs or eVTOLs and their relative motion with respect to BSs. The third cornerstone is system-level communication and network-level handover dynamics, which includes time-varying interference statistics, spectrum sharing, scheduled handover, and their coupling with delayed or quantized CSI feedback. All these cornerstones form a closed-loop simulation environment that reproduces the key characteristics of physical low-altitude FA networks.

\begin{figure*}
    \centering
    \includegraphics[width=0.9\linewidth]{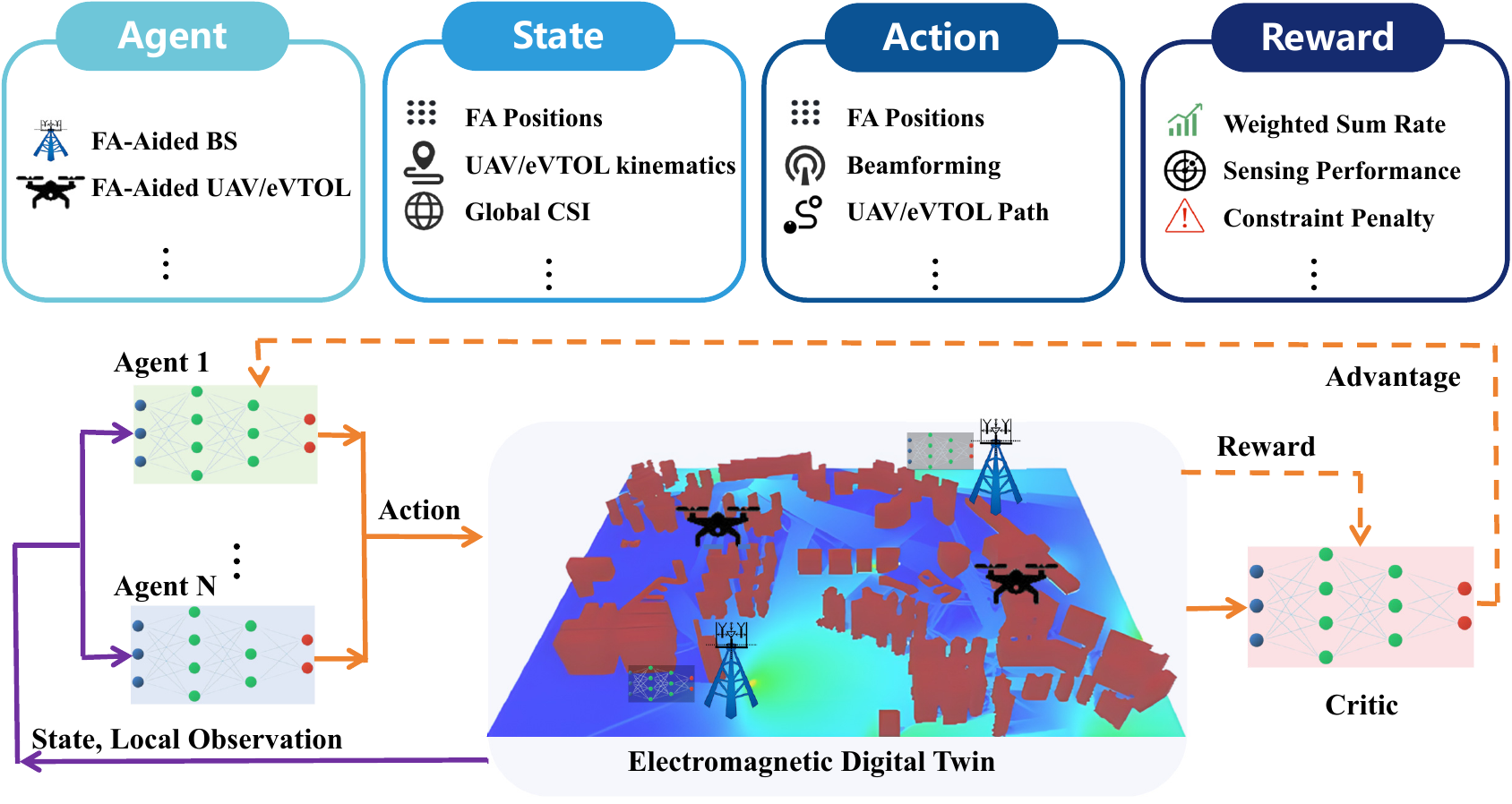}
    \caption{Illustration of MARL training with the EM-DT, and typical Dec-POMDP component definitions.}
    \label{F3}
\end{figure*}

EM-DT can be used in three complementary ways. First, EM-DT provides a low-cost, risk-free, and highly parallel training environment for MARL algorithms. Second, EM-DT serves as a testbed for validating candidate policies under sudden blockage, trajectory deviations, or interference bursts before deployment. Third, EM-DT supports sim-to-real transfer learning through domain randomization over propagation, interference, and mobility parameters. Compared with purely measurement-driven adaptation, which reacts after pilots and feedback reveal wireless channel changes, EM-DT can predict near-future radio maps from mobility and environment priors, reduce measurement overhead, improve sample efficiency, and expose rare safety-critical events. Thus, with sufficient calibration, the EM-DT acts as both a training accelerator and a means of reducing the sim-to-real gap.

\subsection{Multi-Agent Reinforcement Learning}

MARL offers two irreplaceable benefits. The first benefit is its ability to progress policies to high-performance solutions via iterative environment interaction and actor-critic learning. The second benefit is the millisecond inference of trained policies. Mathematically, we formulate the low-altitude FA network optimization problem as a decentralized partially observable Markov decision process (Dec-POMDP), which can be tailored to different task requirements. We illustrate the MARL training with the EM-DT, and typical agent, state, action, reward definitions in Fig. \ref{F3}.

\textit{1) State:} Depending on the specific tasks, the state may include FA positions, UAVs and eVTOLs kinematics, CSI, interference statistics, and other relevant indicators. The state should provide the critic network with sufficient information to enable cooperative credit assignment across agents, while avoiding redundant and  highly correlated features that would enlarge the input space and increase sample complexity.

\textit{2) Local Observation:} Depending on the specific task, the local observation is the information independently received by each agent from the environment. It may include the agent's own FA positions, kinematics, CSI, interference conditions, and other relevant indicators. Local observations complement the global state during centralized training and drive the actor network during decentralized execution.

\textit{3) Action:} Depending on the specific task, the action corresponds to decision variables. 
It may include FA positions, beamforming vectors, power allocation, UAV and eVTOL path adjustments, and other controllable variables. Each agent selects an action based on its local observation to interact with the environment. In real systems, FA position actions should be constrained by hardware speed, movement latency, and settling time, so an agent should update continuous FA positions within a feasible moving region instead of scanning the entire position space in every coherence interval.

\textit{4) Reward:} Depending on the specific task, after all agents execute their actions, the environment will return a reward. The reward is typically defined as a network utility, e.g., weighted sum rate or a weighted combination of sum rate and sensing performance. Constraints can be enforced via adding penalty terms. For tasks with specific goals that lead to sparse rewards, reward shaping can be applied.

Based on the above Dec-POMDP formulation, MARL training and execution can follow the centralized training with decentralized execution (CTDE) paradigm \cite{lowe2017multi}. During training, a centralized critic accesses global states from the EM-DT, including all agents' observations, actions, and environmental information. During execution, each agent acts solely on its local observation using its own actor network, without communicating with other agents. This training--execution separation makes CTDE particularly suitable for low-altitude FA networks: centralized access to global EM-DT information stabilizes cooperative credit assignment and improves training efficiency, whereas communication-free decentralized execution enables millisecond-timescale FA reconfiguration under rapidly varying low-altitude channels.



\begin{figure*}[t]
\centering
\begin{subfigure}{\textwidth}
  \centering
  \includegraphics[width=\linewidth]{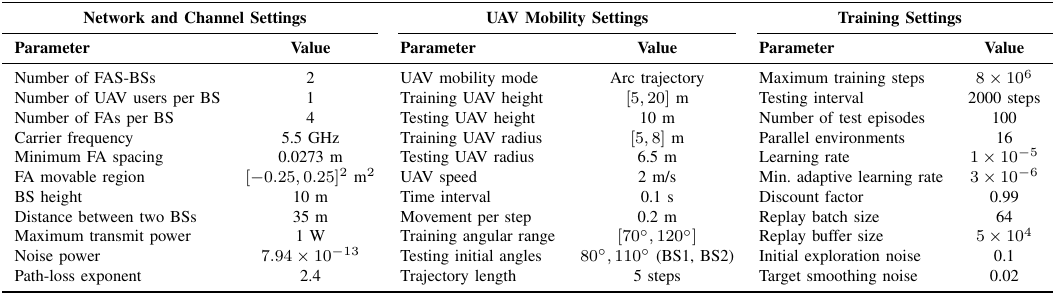}
  \caption{Simulation parameters.}
  \label{para}
\end{subfigure} \vspace{0.01cm} \\
\begin{subfigure}{0.245\textwidth}
  \centering
  \samefig{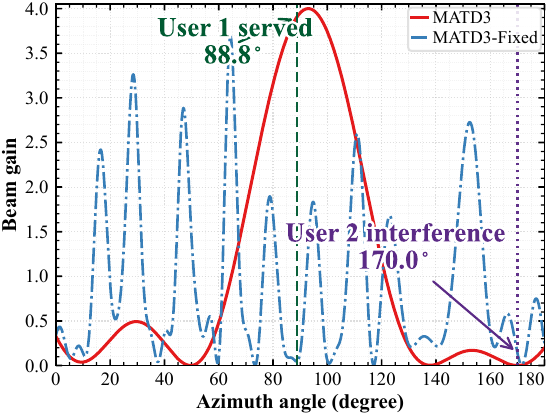}
  \caption{Beampattern at FAS-BS 1.}
  \label{fig:bs1_beampattern}
\end{subfigure}
\begin{subfigure}{0.245\textwidth}
  \centering
  \samefig{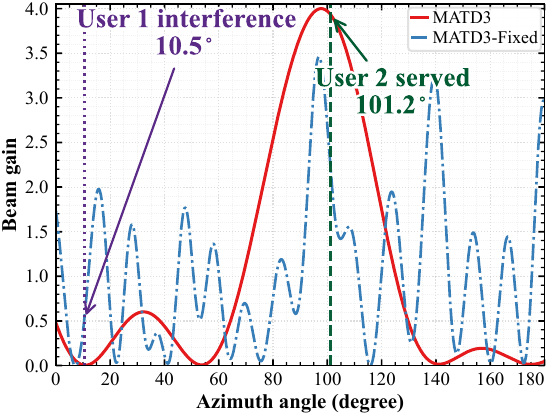}
  \caption{Beampattern at FAS-BS 2.}
  \label{fig:bs2_beampattern}
\end{subfigure}
\begin{subfigure}{0.245\textwidth}
  \centering
  \samefig{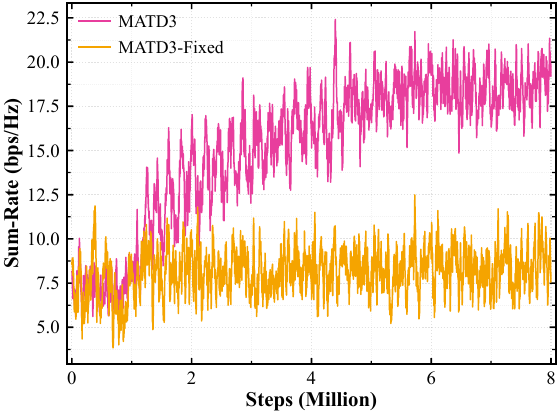}
  \caption{Training curve.}
  \label{fig:training}
\end{subfigure}
\begin{subfigure}{0.245\textwidth}
  \centering
  \samefig{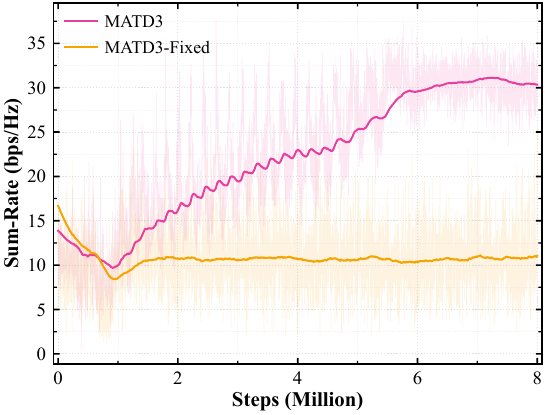}
  \caption{Testing curve.}
  \label{fig:testing}
\end{subfigure}
\caption{Case study of low-altitude FA networks with EM-DT-assisted MARL. Using FAs achieves $118.5\%$ network sum-rate gain under a mobility and unseen environment during testing.}
\label{fig:performance}
\end{figure*}

\section{Sim-to-Real Transfer Learning}

In low-altitude FA networks, the sim-to-real gap could be significant, due to the unique characteristics of FAs and low-altitude airspace, as well as the difficulty of accurate EM-DT modeling. On one hand, real system hardware impairments, FA switching latency, mutual coupling effects are difficult to model accurately in EM-DTs, yet they directly impact the system utility and even the feasibility of the MARL policy's actions. On the other hand, real system factors such as dynamic building blockages, UAV and eVTOL dynamics, weather-induced attenuation, and global position system (GPS)-denied conditions may introduce channel variations that cannot be fully captured during offline training. Furthermore, directly training MARL policies on real low-altitude FA networks would require extensive real system trial-and-error interactions that violate safety guarantees, exceed acceptable convergence time, and risk damaging physical hardware. Therefore, we need transfer learning, which allows MARL policies adapted to the real system with limited measurements \cite{wu2023humanguided}.

We perform large-scale pre-training in an EM-DT, leveraging offline and off-policy MARL algorithms like multi-agent twin delayed deep deterministic policy gradient (MATD3) \cite{ackermann2019reducing}. A key advantage of offline and off-policy MARL algorithms over online, on-policy ones is their superior sample efficiency, achieved through dataset reuse. This flexibility also eliminates the requirement that data be produced by the current policy, allowing for the use of experimentally generated datasets. The introduced transfer learning framework comprises two stages. 

In the Stage-I, multiple agents are trained on pre-collected datasets sourced from the EM-DT. The core objective is to exploit the EM-DT's low cost and high safety, enabling offline and off-policy MARL methods to learn a robust decision model from tens of thousands of pre-sampled trajectories. Most importantly, domain randomization techniques are incorporated during this stage to cover various factors that may appear in real systems, including traffic fluctuations, CSI errors, dynamic topology, and etc.

In the Stage-II, pre-trained offline and off-policy MARL policies can initialize transfer learning in real systems. To mitigate action distribution shifts caused by the absence of real-world feedback, the agents are initially trained on high-quality datasets manually collected from the actual system. These pre-collected datasets are then leveraged in a centralized training phase to update the policies, fully embodying the off-policy paradigm where data acquisition remains decoupled from policy execution.

\section{Case Study}

The simulation serves to preliminarily validate the effectiveness of the proposed EM-DT-assisted MARL framework in improving the downlink performance of LAWNs under aerial-user mobility and time-varying wireless channels. The EM-DT consists of 2 FAS-BSs in free-space airspace, each serving $1$ UAV and equipped with $4$ FAs.  Under the maximum transmit power constraint, the feasible FA moving region, and the minimum FA-spacing limitation, each FAS-BS jointly optimizes the FA positions and downlink beamforming to maximize the network sum-rate. During training, the critic network can access the state with global information, while during execution, each BS makes decisions based only on its local observations with the millisecond inference speed \cite{mxnet2023inference}. The FA positions should be optimized within a bounded two-dimensional (2D) region, subject to a minimum FA spacing constraint. While, this optimization problem is high-dimensional and non-convex. We therefore formulate this optimization problem as a Dec-POMDP and solved by MATD3 under CTDE paradigm. Only the Stage-I in Section IV  is considered in this simulation. Thus, all MARL training and testing phases are executed in an EM-DT.  The other simulation parameters are provided in  Figs.~\ref{fig:performance} (\subref{para}). Code is available at \url{https://github.com/yanfeisu/COM_MATD3}.

\newcommand{\tcell}[2][7mm]{\parbox[c][#1][c]{\linewidth}{\centering #2}}

\begin{table*}[t]
\centering
\caption{Summary of open issues and future directions.}
\label{tab:summary}
\footnotesize
\renewcommand{\arraystretch}{0.95}
\setlength{\tabcolsep}{2.8pt}

\begin{tabularx}{\textwidth}{
>{\centering\arraybackslash}p{0.07\textwidth}
>{\centering\arraybackslash}X
>{\centering\arraybackslash}X
>{\centering\arraybackslash}X
>{\centering\arraybackslash}X
>{\centering\arraybackslash}X
>{\centering\arraybackslash}X
}
\toprule
&
\makecell{\textbf{FA}\\\textbf{Reconfiguration}\\\textbf{Overhead}}
&
\makecell{\textbf{CSI Usage}\\\textbf{Reduction}}
&
\makecell{\textbf{Trustworthy}\\\textbf{MARL}}
&
\makecell{\textbf{EM-DT}\\\textbf{Fidelity}}
&
\makecell{\textbf{Heterogeneous}\\\textbf{Time-Scale}}
&
\makecell{\textbf{System Support}\\\textbf{for Artificial}\\\textbf{Intelligence}} \\
\midrule
\makecell{\textbf{Open}\\\textbf{Issues}}
&
\makecell{Reconfiguration\\cost \& budget}
&
\makecell{CSI overhead\\CSI aging}
&
\makecell{Black-box policy\\Certification\\barriers}
&
\makecell{Static maps\\Dynamic blockages}
&
\makecell{Different dynamics\\Time-scale mismatch}
&
\makecell{Heterogeneous\\data values} \\
\makecell{\textbf{Future}\\\textbf{Directions}}
&
\makecell{Predictive positioning\\\& hardware\\co-design}
&
\makecell{CSI prediction\\by EM-DT}
&
\makecell{Safe MARL\\Certification\\testbeds}
&
\makecell{Surrogate models\\Online calibration}
&
\makecell{Hierarchical MARL\\Asynchronous\\training}
&
\makecell{Task-oriented\\end-to-end design} \\
\bottomrule
\end{tabularx}
\end{table*}

Figs.~\ref{fig:performance} (\subref{fig:bs1_beampattern}) and (\subref{fig:bs2_beampattern}) illustrate the beampatterns of the 2 FAS-BSs. In contrast to the fixed antenna position baseline, the FAS-BSs can adaptively reconfigure the positions of their FAs after MATD3 training, thereby steering the main lobes toward the desired aerial users while effectively suppressing interference directed at undesired users. Fig.~\ref{fig:performance} (\subref{fig:training}) presents the training performance curve. Despite noticeable fluctuations in the training curve, attributed to the randomized UAV heights, trajectories, and channel realizations,  the joint optimization of FA positions and beamforming consistently outperforms the fixed-position baseline in terms of the network sum rate. Fig.~\ref{fig:performance} (\subref{fig:testing}) evaluates the testing performance under fixed UAV height and trajectory conditions that were not encountered during training. In Fig.~\ref{fig:performance} (\subref{fig:testing}), the solid curve represents the exponentially moving-averaged performance with a smoothing factor of $0.99$. The moving-averaged curve in Fig.~\ref{fig:performance} (\subref{fig:testing}) reveals that MATD3 achieves a $118.5\%$ higher network sum-rate compared to the fixed antenna position baseline, even when evaluated under previously unseen conditions.

\section{Open Issues and Future Directions}

Despite the promising potential of low-altitude FA networks, several fundamental challenges remain unresolved. In this section, we identify key open issues that hinder practical deployment and outline corresponding future research directions, as briefly summarized in Table~\ref{tab:summary}.

 

\subsection{FA Reconfiguration Overhead}

Practical FAS faces non-negligible switching delays and hardware constraints. In fast-moving UAV/eVTOL scenarios with speeds up to 30~m/s, the coherence times at high carrier frequencies can range from below one millisecond to several milliseconds. At 5.5~GHz, for example, a 30~m/s velocity induces a Doppler shift of ~550~Hz, translating to a coherence time on the order of 1~ms. Given microsecond-level RF switching, millisecond-level actuator motion, and additional calibration/settling overhead, the practical reconfiguration rate is inherently limited. Repeated FA scanning and physical repositioning for channel estimation may thus exceed the coherence budget, erasing theoretical benefits. This calls for hardware-aware control strategies that consider switching costs, per-coherence-interval budgets, and predictive positioning via temporal channel correlation.

\subsection{CSI Usage Reduction}
MARL frameworks rely on perfect or instantaneous CSI during deployment. In practice, however, accurate CSI acquisition comes with high overhead from pilot transmissions, estimation, and feedback. Additionally, in fast-varying environments, the acquired CSI often becomes outdated before use. This calls for alternative solutions, including domain randomization and meta-learning for robustness against CSI imperfections, partial observation for inferring channel dynamics from historical data, event-triggered CSI acquisition, and EM-DT with channel prediction to replace direct measurement.

\begin{figure*}[t]
    \centering
    \includegraphics[width=0.96\textwidth]{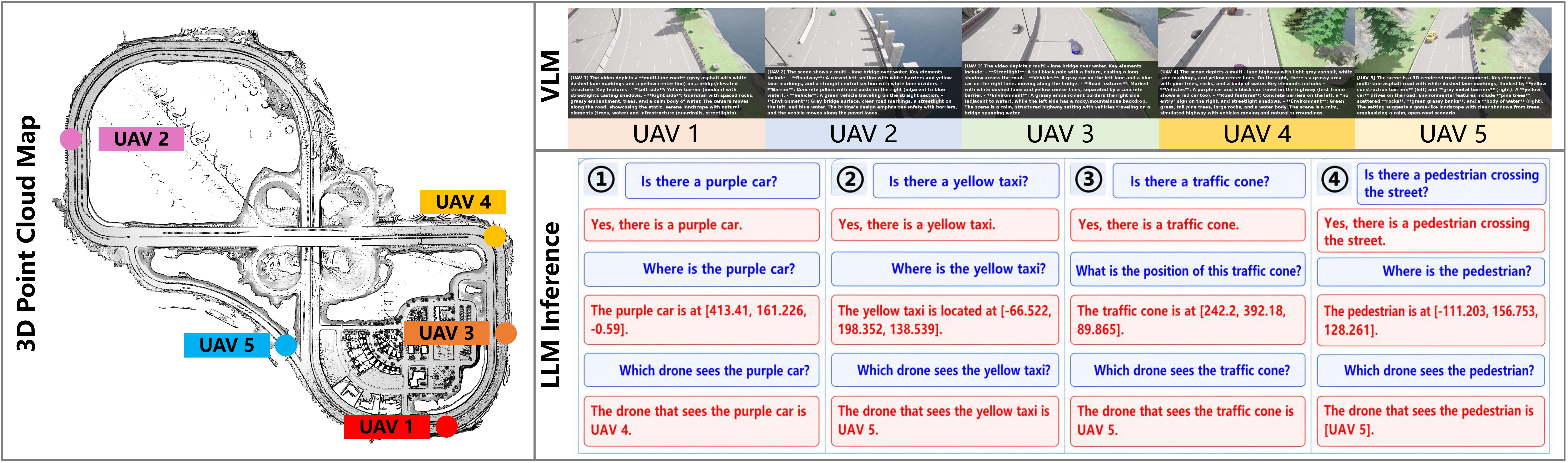}
    \caption{Low-altitude FA networks for aerial visual question answering.}
    \label{fig:use}
\end{figure*}

\subsection{Trustworthy MARL}
Low-altitude airspace hosts safety-critical missions, such as eVTOL flights and emergency response, which demand FA controllers that are both effective and reliable. Yet MARL policies are data-driven and stochastic by nature, lacking worst-case performance guarantees. Their decision-making is opaque and hard to verify, and unexpected issues may arise. There is thus an urgent need for safe MARL integration, certification frameworks, and standardized testbeds. 

\subsection{EM-DT Fidelity and Calibration}

The fidelity of an EM-DT critically depends on the accuracy of its underlying environmental maps and propagation models. Current EM-DTs rely on static or slowly updated maps of building geometries and material properties, yet real low-altitude airspace is characterized by dynamic obstructions such as moving UAVs, vehicles, and weather conditions. Such map inaccuracies directly degrade the fidelity of ray-tracing predictions. Moreover, 3D ray-tracing alone may not match the accuracy of real measurements. To maintain high fidelity, the EM-DT must be continuously calibrated by assimilating sparse real-system measurements. This calls for a hybrid approach that combines 3D ray-tracing with lightweight, data-driven surrogate models, enabling the EM-DT to efficiently incorporate sparse measurements and maintain alignment with the evolving physical environment.

\subsection{Decision Making in Heterogeneous Time-Scale}

Low-altitude FA networks face a significant challenge from heterogeneous dynamics and temporal scale mismatch between BSs and aerial UAVs or eVTOLs. BSs are static, whereas UAVs and eVTOLs move at high speeds, forcing the MARL policies to process state information across disparate scales. Moreover, during decentralized execution, agents may operate at inconsistent observation and decision-making frequencies. Hence, hierarchical MARL frameworks are urgently needed, in which slow-timescale variables manage long-term resource allocation and handover, fast-timescale variables handle instantaneous FA reconfiguration and beamforming, and asynchronous training with communication mechanisms coordinates decisions across temporal scales.

\subsection{System Support for Artificial Intelligence}

Low-altitude FA networks are envisioned to enable diverse artificial intelligence (AI) applications. For instance, in aerial visual question answering, users may query UAV teams regarding observations over a long horizon. Captured images are uploaded via FA networks to a ground server, where vision-language models (VLMs) generate textual descriptions and large language models (LLMs) answer queries via in-context learning.
However, UAVs retain heterogeneous semantic information due to varying observations. As shown in Fig.~\ref{fig:use}, while UAVs 1–3 observe no notable events, critical semantic cues regarding the presence and positions of the purple car, taxi, traffic cone, and pedestrian are derived solely from UAVs 4–5. This underscores the necessity for task-cognitive FA networks capable of intelligently fusing heterogeneous knowledge to optimize end-to-end task performance.

\section{Conclusion}

In this paper, we investigated applying FAs to overcome the limitations of traditional MIMO in low-altitude wireless networks with highly dynamic aerial environments. We proposed an EM-DT-assisted MARL framework that enabled safe and efficient offline training of decentralized policies, together with a transfer learning mechanism that bridged the sim-to-real gap. Simulation results on a two-BS two-UAV case study demonstrated that joint optimization of FA positions and downlink beamforming consistently improved the sum-rate over the fixed-position baseline by dynamically reshaping beampatterns toward UAVs. We also identified key open issues and
outline promising directions for future research.

\balance

\begin{IEEEbiographynophoto}{Tong Zhang} [M] (tongzhang@hit.edu.cn) received the B.S. degree from Northwest University, Xi’an, China, in 2012, the M.S. degree from Beijing University of Posts and Telecommunications, Beijing, China, in 2015, and the Ph.D. degree in electronic engineering from The Chinese University of Hong Kong, Hong Kong, in 2020. He was a Post-Doctoral Fellow with the Southern University of Science and Technology from 2020 to 2022. He was a Lecturer with the Department of Electronic Engineering, Jinan University. He is currently an Assistant Professor with Harbin Institute of Technology, Shenzhen. His research interests include fluid antenna systems and reinforcement learning.
\end{IEEEbiographynophoto}

\begin{IEEEbiographynophoto}{Yanfei Su} (2022210389@stu.hit.edu.cn) received the B.S. degree in communication engineering from Harbin Institute of Technology (HIT), Weihai, China. She is currently pursuing the M.S. degree in information and communication engineering with HIT, Shenzhen, China. Her research interests include fluid antenna systems and reinforcement learning.
\end{IEEEbiographynophoto}

\begin{IEEEbiographynophoto}{Shuai Wang} [M] (s.wang@siat.ac.cn) received the B.Eng. and M. Eng. degrees from Beijing University of Posts and Telecommunications ( BUPT), in 2011 and 2014 respectively, and the Ph.D. degree from University of Hong Kong (HKU) in 2018. He is now a Professor with the Shenzhen Institutes of Advanced Technology (SIAT), Chinese Academy of Sciences, where he leads the Intelligent Networked Vehicle Systems (INVS) Laboratory. His research intrests include robot learning and networked intelligence.
\end{IEEEbiographynophoto}

\begin{IEEEbiographynophoto}{Wanli Ni} [M] (niwanli@bupt.edu.cn) is currently an Assistant Professor with the School of Information and Communication Engineering, Beijing University of Posts and Telecommunications (BUPT), China. He received the B.Eng. degree in 2018 and the Ph.D. degree in 2023, both from BUPT. From 2022 to 2023, he was a visiting student at the Nanyang Technological University, Singapore. From 2023 to 2025, he was a Postdoctoral Researcher at the Tsinghua University, China. His research interests include federated learning, large AI models, spectrum sensing, and satellite communications.
\end{IEEEbiographynophoto}

\begin{IEEEbiographynophoto}{Chengzhong Xu} [F] (czxu@um.edu.mo) received the Ph.D. degree from the University of Hong Kong, Hong Kong, in 1993. He has held Faculty positions with Wayne State University, Detroit, MI, USA, and the Shenzhen Institutes of Advanced Technology, Chinese Academy of Sciences, Shenzhen, China. He is currently the Chair Professor of computer science with the University of Macau, Macau, China. He is also the Dean of Faculty of Information Science and Computing and the Director of Institute of AI and Brain Sciences. His research interests include cloud and edge for AI, autonomous driving, and intelligent transportation.
\end{IEEEbiographynophoto}

\begin{IEEEbiographynophoto}{Hüseyin Arslan} [F] (huseyinarslan@medipol.edu.tr) is currently the Dean of the School of Engineering and Natural Sciences at Istanbul Medipol University, Türkiye. He received his bachelor’s degree from Middle East Technical University, Türkiye, in 1992, and his M.S. and Ph.D. degrees from Southern Methodist University, USA, in 1994 and 1998, respectively. From 1998 to 2002, he was with Ericsson Research, where he worked on 2G and 3G wireless communication systems. From 2002 to 2022, he was with the University of South Florida, USA, where he served as a Professor of Electrical Engineering. In 2013, he joined Istanbul Medipol University and contributed to the establishment of its engineering school. He has also held various academic, industrial, and advisory roles and has served on the editorial boards of several IEEE journals. His research interests include 6G and beyond wireless communications, waveform design, spectrum sharing, physical-layer security, interference management, non-terrestrial networks, and integrated sensing and communication.
\end{IEEEbiographynophoto}

\end{document}